# Virtual learning: possibilities and realization
*Kerimbayev Nurassyl*
*Kazakhstan, Astana*

**Abstract.**
The virtual learning in High University Education is the learning which is presented by set of integrated information and pedagogical technologies, in a process of interaction between subjects and objects as the virtual educational resources. This interaction characterize as the set of dialectically interconnected fields of human activity (intellectual, emotional and figurative, cultural, social). The virtual educational resources, possibility of their adaptation to student subjectivity, and realization in the conditions of High University education are the main issues of this article.

**The main issue.**
We consider virtual learning as a component of virtual education. Thus virtual education is understood as the educational environment, virtual space in which subjects and objects of educational process interact by electronic communications tools.

## 1. Introduction

In the modern world the concept of virtuality is considered and studied in philosophy, psychology, physics, biotechnology, the sphere of arts, etc. Specific properties of virtual reality are relevance, autonomy, interactivity. It is possible to allocate the following virtual educational process keys:
• preliminary uncertainty for subjects' interaction;
• uniqueness for each sort of interaction, including interaction with real educational objects;
• existence only throughout direct interaction.

The term "virtual" from the Latin word "virtualis" means "the possible; which can or has to appear under the certain conditions" [1]. Any virtual process happens in the corresponding virtual space which properties are defined by similar signs and virtual objects existence in it. Therefore the virtual educational space demands educational process objects and subjects existence as the virtual learners and virtual teachers. Virtual learning acts as the process which is directed on result. Thus, Virtual learning is understood as process and result of subjects and the objects interaction. This interaction defines educational process specifics as a whole system.

Educational process is as virtual as there are the subject and the object of relations. Virtualization of this process is the education virtualization.

The virtual educational environment as set of information resources provides complex methodical and technological support for educational process, educational process management, and its quality. Therefore, it is possible to say that the virtual educational environment carries out the following functions: information training, communication, administrative function. The virtual educational environment is a virtual interaction between educational process participants with help of information and communication technologies. The information and educational environment is the general information space and the virtual educational environment is its part. For the virtual educational environment the common is the Virtual learning environment existence. This environment represents as complex computer tools and technologies which allow controlling the educational environment maintenance and educational process participants' communication [2].

In this article virtual educational environment represents as multipurpose system where specific pedagogical, didactic and methodical technologies are realized, and information resources (a database and knowledge, library, electronic training materials), modern software are necessary.

Virtual education is somehow similar to distance learning. However, subjects and objects interaction happens in real time, interaction between teachers, student and studied objects. Thus it has to be considered that the teacher can act indirectly, i.e. the teacher role can be replaced by digital technology, computer program, etc.

Virtual learning is a process and the educational process participants' communication results are in the virtual environment.

Virtual learning is distinguished by the following advantages: training environment mobility and interactivity, distantness, information educational resources existence.

The virtual education purpose is the place and achievement identification by the person in real world, including its virtual component.

At Virtual learning the multimedia educational production is realized as the digital educational resources. Digital educational resources act as objects of virtual reality and interactive modeling, cartographic materials, sound recordings, symbolical objects and business graphics, text documents and other training materials which are necessary for the educational process organization. The digital educational resource is the information object. If consider that Virtual learning has informative character, the learner's main problem is not only access to information, but also to know can presented information usage, which is not always structured and organized, and often can have unsatisfactory characteristics.

This article is urged to open opportunities and to plan ways of Virtual learning realization in High University education. The main problem we faced in this article is the virtual educational space organization issues. This space has to carry out organizational focusing functions, and directed on remote education realization, auxiliary electronic resources creation for supplementing databases for High university education.

**II. Methodology**

To define bases of virtual learning, it is necessary to consider theoretical and methodological aspects of virtual reality. First of all the research of this article demands to establish philosophical basis for pedagogical and didactic development of virtual learning bases.

Philosophical aspects of virtuality are known from medieval culture. Philosophers of that time explain the term "virtus" as a soul, "as the active beginning inherent in object internal ability, a potentiality which is realizing not in reality under the corresponding conditions" (Cicero, F.Akvinsky, S. Brabansky). Psychologist A.N. Leontyev said that "virtually brain comprises not these or those specific human abilities, but these abilities formation". The virtual reality – is special philosophical category such as time, space, essence, etc. These categories relate to the different types: Natural Science, Humanitarian or Technical". According to the people existence philosophical concepts the potentially existing realities have virtual status. Psychological virtual realities are generated by person's mentality. The virtual reality exists while the reality generating is active; and virtual reality has own time, space and laws. The virtual reality can interact with all other realities. [3]

Virtual reality creation technologies are used in computer games, exhibitions and commercial representations, in space exercise machines and other kinds of activity where modeling of real and virtual processes is required. Prompt development of information, human telecommunications and technologies, the global Internet, creation of the various devices providing people interaction with virtual reality (3D - points, 3D - helmets, technologies 5d, etc.). These technologies are generating virtuality - the complex scientific subject which is studying virtuality and virtual reality problems. Virtuality gives the chance for computer virtual realities technology adequate inclusion in all spheres of human life: education, medicine, biotechnologies, etc. For instance, E. Blinnel developed virtual reality training platform as Fraunhofer IFF Learning a platform. [4]

As the results of virtual learning (as any other type of learning) student receives the necessary sum of knowledge, abilities, and gains professional competences. However, these results are not defined as forms of educational activity, but as basic philosophical meanings, building educational process.

Virtual reality from philosophical aspect point of view is important part of virtual training. Only in this case it is possible to speak about cultural aspect of education system development. For virtual education philosophical model creation the methodology is defined by cognizable object, instead of who learns. This conclusion contradicts to the modern school education development idea. Any object is capable to bring the person for understanding the internal meanings of life. The conclusion define that there is no specific education scheme.

Theoretical and philosophical framework of virtual learning is determined by Levy (2000), Zanine (2003), Rosa (2004). [5] For virtual learning number of the general principles and practical methods, which can be significant for virtual education preconditions identification, is important: the relation between person and the. The content of education has gradual spatial expansion.

First of all, virtual education is a movement from new to unknown, while interaction with the real world. Student's virtual education is expansion of his or her world from inner to external level, and interprets as micro and macro mass.

Virtual education spatial model creation represents as set of extending spheres: intellectual, emotional and figurative, cultural, historical, social and others. All of them are closely connected, and can be define as students' virtual educational space.

### III.I Virtual education in terms of future students' Professional Development system

Nowadays virtual education took an important place in future students' professional development system, and during rapid growth of scientific and technical potential the need for the highly qualified specialists, perfectly knowing the subject, is growing rapidly too.

The aim of virtual education – identification and achievement by the person of the mission in the real world combined with its virtual and other opportunities. Virtual education for future students' professional development system has a subject focus, and represents person's "open" virtual university (J.F. Lapointe, H. MacLean, K. Popat).

Xinghong Liu had studied the virtual education impact for future students' professional development importance issue. This research based on virtual educational communities' experience: virtual learning communities or online communities are used by a variety of social groups interacting via the Internet. Different virtual communities, like real communities, have different levels of interaction and participation among their members. An important characteristic of a community is the interaction among its members. [6]

Virtual education for student gives interactive access to the digital libraries provides with powerful search engines, can study at home or where there is an Internet access.

Virtual educational space modeling for higher university education based on of competence-based approach for students professional development is actual than ever. One way or virtual education realization is virtual reality technology application where high-realistic professional activity multicomponent space modeling supports dynamic interaction with students. In this process the complex psychology-pedagogical conditions, where the virtual education opportunities realization are possible have to be implemented.

Virtual educational space creation for high university education is innovative in future students' professional development, and directs on professional education quality improvement.

Virtual educational space creation for high university education has to take into account properties of virtual reality, and based on virtual outlook phenomenon.

The virtual educational environment is realized in the conditions of pedagogical interaction in space and time, online. These signs provide various models creation optimum conditions.

Virtual reality forms new forms of mentality and consciousness activity manifestation, and the return impact on factors which have generated them, and human life. The virtual educational environment influences to human knowledge and reality transformation; on human activity, worthiness, self-determination, and self-realization. Therefore the virtual educational environment has to be based on recognition of humanistic essence and focused on professional education.

The virtual educational environment of high university education consists of these main components: informative, integrative, communication, coordinating, developing and professional focusing. These cognitive and logic-subject components provide professional knowledge individual conceptual system formation.

The virtual educational environment organization for high university education the educational models, methods and receptions future students professional development have to be focused on virtual education opportunities realization and provide training process reflexive management.

The virtual educational environment as part of system of a common information space allows increasing the students' professionalism.

### III.II Virtual learning possibilities

What virtual education possibilities and what advantages of training do we use of virtual technologies? First of all it is necessary to allocate the psychologist – pedagogical opportunities which provide virtual images generation possibility, develop theoretical, intuitive, creative virtual thinking.

Virtual learning possibility based on functional orientation provides the systematic interdisciplinary Virtual learning; the resources providing possibility for formal and logical, cause and effect reflections, objects functioning with use of digital educational resources in terms of virtual educational environment.

Virtual learning is realized such kinds of activity, as poly-subject and mono-subject. Poly-subject activity is focused on personal model of interaction and provides opportunity to carry out communicative interaction. Mono-subject activity is focused on self-development and self-realization and provides possibility for students' independent work optimization within their interaction with electronic educational resources.

Efficiency of virtual education realization can be estimated by the following criteria:
• level of motivational readiness formation;
• the organization of complex interdisciplinary Virtual learning based on approach;
• information and technological skills development degree for virtual education realization;
• students' creativity and independence in terms of virtual education realization.

### IV. Technology of virtual learning realization opportunities.

Process of virtual education realization opportunities for future students' professional development is based on differentiated, individual, procedural, competence-based approaches.

The technology of virtual education realization opportunities for students' professional development unites the following set of interconnected and interdependent structural components: classification of virtual education opportunities; levels and stages of virtual education realization opportunities.

Nowadays education virtualization represents set of internal, remote education and self-education as the result occur the communication network (on-line) mobility, multimedia and telecommunication systems development. For instance the "Virtual learning" portal is created for L.N. Gumilyov the Eurasian National University of Astana.

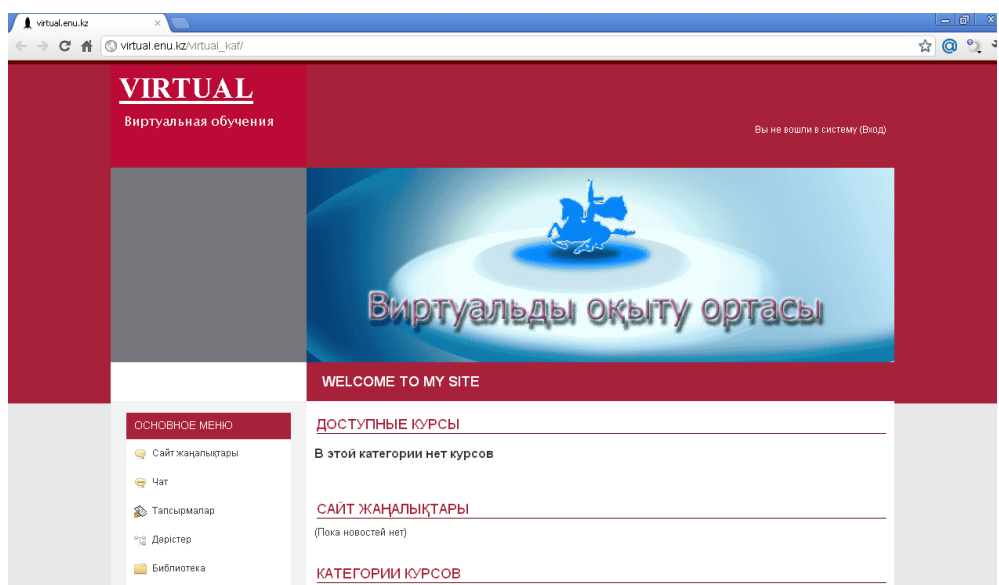
Image 1

This portal gives us the search and access possibilities for all necessary information we need. There are training programs, glossary, articles, books (scanned for full text reading), and information about analytical, scientific and methodical centers. The portal structure is for interactive informational content.

This portal created for educational resources granting possibility for being "distance" learning. Moreover, it opens additional elective disciplines studying possibility from the High University curriculums which contain in Virtual Center, from author's original courses, and training materials. The "Virtual learning" portal provides the developed virtual infrastructure, including electronic library, participation in electronic conferences available in real time.

The "Virtual learning" portal is the tool and provide reliable access to the interesting for users different contents, applications, and the services organized as a unit. Furthermore, it gives opportunity for content choice conforming to educational standards requirements. The information content is based on learners' creative, moral, intellectual qualities, and humanity development.

The "Virtual learning" portal creation was a logical stage of the L.N. Gumilyov the Eurasian National University's virtual educational space organization.

**V Conclusion.**

Virtual learning is the system of the components presented by set of integrated information and pedagogical technologies, and interaction between subject and object as virtual educational resources.

Virtual training realization opportunities provide mass training opportunities, modularity and methodical toolkits unification, standardization, developed possibility, and its high-quality and quantitative updating.

The main components of virtual learning are directed on realization of the pedagogical principles and modern training process technologies and experts future development:

| | | |
|---|---|---|
| Main leading idea | ⇒ | students professional development efficiency and intensification on levels diversification, variability, differentiation and a training individualization; |
| Attributes | ⇒ | virtual education openness, virtual educational space discretization, scalability of virtual reality, adaptability of virtual reality technologies, informative learners activity; |
| Tools and software | ⇒ | Special hardware-software interface (Flash Player, Macromedia, Open Meetings, Moodle) and computer virtual reality technologies (WEB technology, VR technology). |

Virtual learning allows high-quality, transparent and available training. The presented virtual educational environment provides virtual learning, management of educational process and the quality control, directed on students' professional development.